\documentclass[aps,prb,twocolumn,showpacs,preprintnumbers]{revtex4-2}
\usepackage{graphicx}
\usepackage{amsmath}
\usepackage{amssymb}
\usepackage{graphicx,epstopdf}
\usepackage{graphicx}
\usepackage{epstopdf}
\usepackage{mathtools}
\usepackage{tikz}
\usepackage{color}
\usepackage[T1]{fontenc}
\usepackage{hyperref}

\usetikzlibrary{shapes.misc,shadows}

\newcommand{\be}{\begin{equation}}
\newcommand{\ee}{\end{equation}}
\newcommand{\bea}{\begin{eqnarray}}
\newcommand{\eea}{\end{eqnarray}}
\newcommand{\bse}{\begin{subequations}}
\newcommand{\ese}{\end{subequations}}

\begin{document}
\title { Inertial active Ornstein-Uhlenbeck particle in a non-linear velocity dependent friction}

\author{N Arsha}
%\author{M Mushin}
\author{M Sahoo}
\email{jolly.iopb@gmail.com}
\affiliation{Department of Physics, University of Kerala, Kariavattom, Thiruvananthapuram-$695581$, India}
\date{\today}
\begin{abstract}
 We explore the self-propulsion of an active Ornstein-Uhlenbeck particle with a non-linear velocity dependent friction. Using analytical approach and numerical simulation, we have exactly investigated the dynamical behaviour of the particle in terms of particle trajectory, position and velocity distribution functions in both underdamped as well as overdamped regimes of the dynamics. Analyzing the distribution functions, we observe that for a confined harmonic particle, with an increase in duration of self-propulsion, the inertial particle prefers to accumulate near the boundary of the confinement rather than the mean position, reflecting an activity induced bistability in the presence of nonlinear friction. On the other hand, in the overdamped or highly viscous regime, where the inertial influence is negligible small, the sharp peak structure in the distribution across the mean position of the well reveals as usual trapping of the particle with increase in the persistent duration of activity. Moreover, for a free particle, using perturbation method, we have analytically computed the velocity distribution function in the vanishing limit of noise. The distribution interestingly shows the similar attributes as in case of a harmonic well, thus providing an additional effective confining mechanism that can be explained as a decreasing function of effective temperature. In this limit, the analytically computed distribution agrees well with the simulation results.
\end{abstract}
\maketitle
\section {\textbf{Introduction}}
Over the past few years, research activities have grown dramatically in the area of statistical description of systems far from equilibrium~\cite{reimann2002brownian,PhysRevLett.117.038103}. Among these systems,  active  matter is one that is inherently driven away from equilibrium. It allows individual units to move actively by gaining energy from the environment~\cite{grossmann2012active, RevModPhys.85.1143, romanczuk2012active}. Some of the distinct features of active matter includes broken detailed balance, broken time reversal symmetry, etc. The behaviour of such systems has been the subject of many theoretical and experimental studies 
~\cite{steuernagel1994elementary,park2013modified,romanovsky2013models,TONER2005170,kumar2014flocking,PhysRevLett.108.235702,PhysRevLett.116.218101}.     
In the recent years, there are lot more progress in the study of active motion both at individual and collective levels of the dynamics~\cite{chaudhuri2014active,romanczuk2012active,hoffmann2002nonlinear}. \\
 At theoretical level, for modelling the active dynamics, one way of driving the system inherently out of equilibrium is to modulate either viscous or noise level of the environment %he activity can be introduced in the dynamics either by modulating the viscous level or the noise level
~\cite{PhysRevE.90.022131,h1989colored,haunggi1994colored,PhysRevE.94.032602,erdmann2000brownian,erdmann2003collective,PhysRevE.87.052135}. The viscous drag can be modulated by considering a more complex friction that may depend on space, velocity or time. Among various models of such active motion, the motion in a non-linear velocity dependent friction helps in pumping energy to the particles from the environment~\cite{romanczuk2012active,erdmann2000brownian}. Different models that use this non-linear velocity dependent friction includes the Rayleigh-Helmholtz (RH) model~\cite{lindner2008diffusion,schimansky2005stationary}, the energy depot model~\cite{ebeling2004nonlinear,ebeling1999active,ebeling2005klimontovich,ebeling1999active,zeng2016impact}, and the Schienbein-Gruler(SG) model~\cite{schienbein1993langevin}. In RH model, up to certain range of (small) velocities of the particle, the friction coefficient can become negative ~\cite{romanczuk2012active} and thus the motion of the particle can be pumped with energy. Whereas in the energy depot model, the particle has the ability to take up energy from the environment, store it in an internal energy depot and convert it into kinetic energy~\cite{erdmann2000brownian}. However, both RH and energy depot model describes a bimodal distribution of velocity in various aspects of motion such as microtubules under the collective influence of the bidirectional motor proteins NK11~\cite{badoual2002bidirectional}, two dimensional motion of a free self propelled particle with internal energy depot~\cite{erdmann2000brownian}, the motion of swarm of interacting self propelled particles with an internal energy depot~\cite{lobaskin2013collective}, etc. 
Similarly, in the last decade, some of the other aspects of modelling active motion 
has been proposed and these includes active Brownian Particle (ABP) model~\cite{hagen2009non,ten2011brownian,PhysRevE.101.022610,lowen2020inertial}, active Ornstein-Uhlenbeck particle (AOUP) model~\cite{PhysRevE.103.032607,PhysRevE.100.022601,PhysRevE.97.012113}, and run and tumble particle (RTP) model~\cite{cates2012diffusive,cates2013active}.
The AOUP model is the simplest one in describing the behaviour of active particles as it makes the exact analytical calculations possible. For the stochastic motion of an Ornstein-Uhlenbeck (OU) particle in the presence of RH friction, the diffusion coefficient shows a minimum as a function of noise correlation time~\cite{lindner2010diffusion}.
Similarly, an OU particle while self-propelling in the presence of a periodic swim velocity exhibits deviations from Gaussian shape in the velocity distribution, even displaying a bimodal behaviour in the high-motility regime~\cite{10.21468/SciPostPhys.13.3.065}.
However, the study of an active OU particle with an internal energy depot has received limited attention.\\ 
In this paper, using both numerical simulation and analytical approach, we investigate the motion of an inertial AOU particle in the presence of an internal energy depot. Both the case of a free particle and a confined harmonic particle are explored by analyzing the particle trajectories and steady state position and velocity distribution functions. For the motion in a confined harmonic well, both position and velocity distribution show two separated sharp peak structure with increase in activity time, confirming a signature of an activity induced bistability. For a free particle, the velocity distribution interestingly shows the similar characteristics, which can be explained by an induced effective confining mechanism and it's  effects can be described as a decreasing function of effective kinetic temperature or noise strength of the medium. Moreover, for low value of noise correlation time, the analytically calculated velocity distribution is in good agreement with the simulation results.

\section{\textbf{Model}}
  We consider the motion of an inertial active Ornstein-Uhlenbeck particle(AOUP) in the presence of a non-linear velocity dependent friction force. The motion can be described by the Langevin's equation of motion~\cite{muhsin2022inertial,arsha2023steady}
 \begin{equation}
      m\frac{dv}{dt}=-\gamma_{v} v-\frac{\partial U\left( x,t\right) }{\partial x} +\sqrt{D}\eta(t), \label{dynamics1}
\end{equation}
where $m$ and $v(t)$ are the mass and instantaneous velocity of the particle. $\gamma_{v} $ is the friction coefficient, which has a non-linear dependence on velocity and is given by
\begin{equation}
    \gamma_{v}=\left(\gamma_{0}-\frac{qd}{c+dv^2}\right).
    \label{dynamics2}
\end{equation}
Here $q$ is the rate at which energy is taken from the environment(pumping rate), $c$ is the rate of energy loss by internal dissipation and $d$ corresponds to the rate of conversion of internal energy into kinetic degrees of freedom. $U\left( x,t\right)$ is the confining potential, which we have considered here as a one dimensional harmonic well ($U(x)=\frac{1}{2}kx^{2}$), with $k$ as the harmonic constant. $\eta(t)$ is basically the noise force, which follows the Ornstein-Uhlenbeck process and satisfies the properties $\langle \eta(t) \rangle=0$ and $\langle \eta(t)\eta(t') \rangle = \dfrac{1}{2t_{c}}e^{-\frac{\vert t-t'\vert}{t_{c}}}$. Here, $t_{c}$ represents the self-propulsion time or activity time of the dynamics and $D$ represents the strength of the noise. The mass of the particle is assumed to be unity ($m=1$) throughout the paper.

\section{\textbf{RESULTS AND DISCUSSION}}
\begin{figure}[ht]
    \centering
\includegraphics[width=0.5\textwidth]{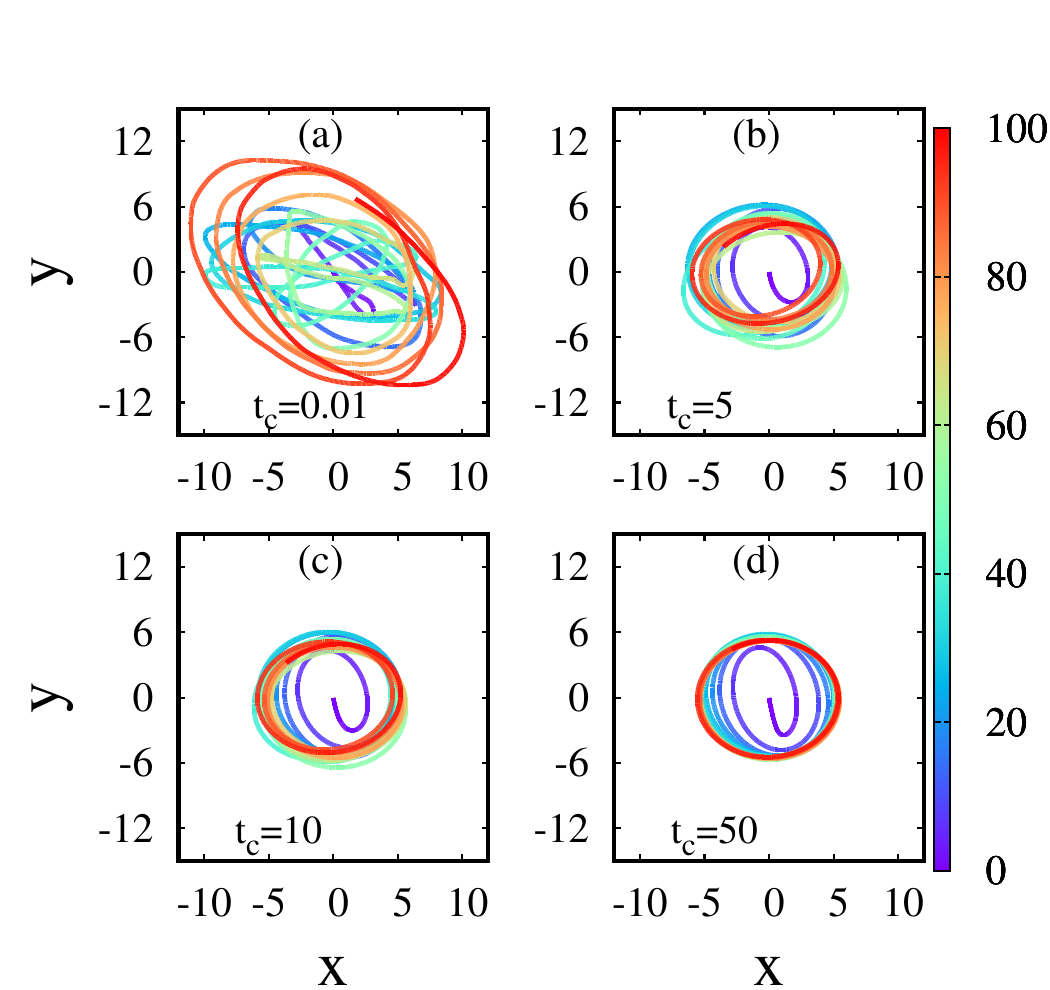}
    \caption{Particle trajectories in $x$-$y$ plane of an inertial active particle for different values of $t_{c}$. The other common parameters are $\text{D}=\text{k}=1, \text{q}=3, \text{c}=0.01, \text{ and }\gamma_{0}=\text{d}=0.1$.}
    \label{Fig1traj}
\end{figure}

We simulated the dynamics [Eq.~\eqref{dynamics1}] using Euler's-Maruyama(EM) method~\cite{kloeden1992stochastic}. The simulation is run for $10^{3}$ time steps with each time step of the order of $10^{-2}$. The steady state averages are taken over $10^{3}$ realizations after ignoring the initial transients of the order of $10^{2}$ time steps. %The steady state position and velocity distribution functions are evaluated using the steady state data for position and velocity.\\
 The simulated particle trajectories of an inertial active particle  in the $x$-$y$ plane for different values of $t_{c}$ is shown in Fig.~\ref{Fig1traj}. For very low $t_{c}$ values or in the vanishing limit of $t_{c}$, the particle shows rotational trajectories as in case of Brownian motion of a particle in nonlinear friction ~\cite{schweitzer1998complex}. For a finite $t_{c}$ value, the particle performs almost elliptical trajectories, which is a typical feature of an inertial Ornstein-Uhlenbeck particle while confined in a harmonic well ~\cite{caprini2021inertial, arsha2023steady}. With increase in $t_{c}$ values, the trajectories get comparatively smoother and the elliptical trajectories are more stabilized. Initially, the trajectories get suppressed with increase in $t_{c}$ values. However, for higher $t_{c}$ values neither suppression nor enhancement of trajectories are noticed. This observation confirms that for sufficiently long duration of activity, the particle don't get confined at the centre of the potential, rather it moves away from the centre and tries to accumulate near the boundary of the potential well.
\begin{figure*}[ht]
    \centering
   \centering
   \includegraphics[scale =0.48]{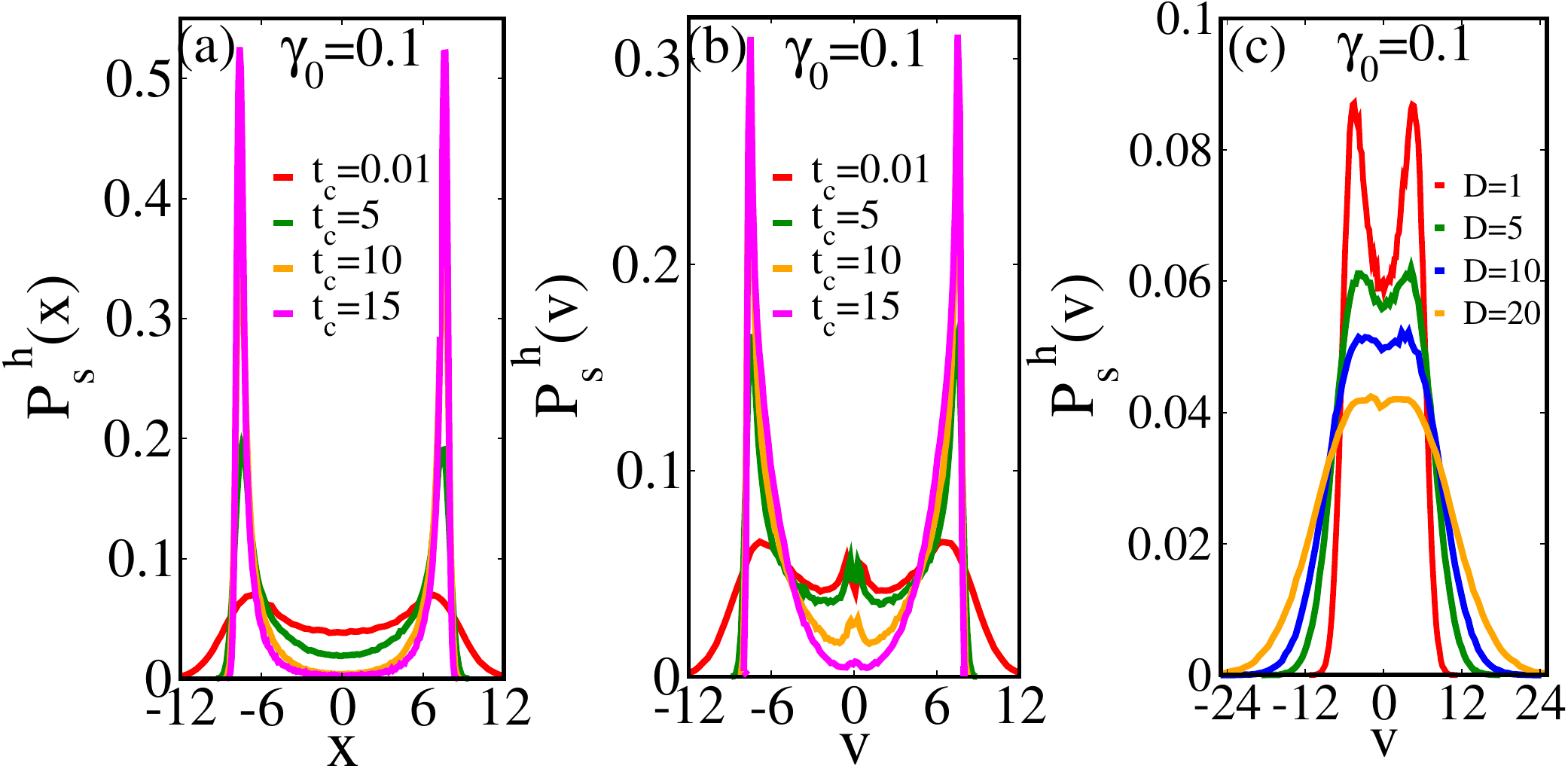}
    \caption{ The position distribution  $P_{s}^{h}(x)$ as a function of $x$ for different values of $t_{c}$ in (a) and velocity distribution $P_{s}^{h}(v)$ as a function of $v$ for different values of $t_{c}$ in (b) and for different values of $D$ in (c), respectively. The other common parameters are $\text{D}=\text{k}=1, \text{q}=3, \text{c}=0.01, \text{ and }\gamma_{0}=\text{d}=0.1$.}
    \label{Fig1}
\end{figure*}\\

 From the simulated data for position and velocity of the particle, we have computed the steady state probability distribution function for position $[P_{s}^{h}(x)]$ and velocity $[P_{s}^{h}(v)]$ of the particle, which is plotted in Fig.~\ref{Fig1}(a) and  Fig.~\ref{Fig1}(b), respectively for different values of $t_{c}$. For very low $t_{c}$ value, the distributions shows a bimodal structure with two peaks, which is a genuine feature of an inertial particle in a nonlinear friction~\cite{erdmann2000brownian}. With increase in $t_{c}$ value, initially the local confinement of the particle gets reduced as reflected from the distribution curve of $[P_{s}^{h}(x)]$ [Fig.~\ref{Fig1}(a)]. However, the particle moves away from the centre of the potential, tries to accumulate near the boundary and spends most of the time there. As a consequence, the probability of finding the particle at the centre of the potential decreases and the probability of finding the particle near the boundary increases. For very high $t_{c}$ values, both position and velocity distribution functions exhibit a transition from bimodal distribution to two separated sharp peak structured distribution across the centre of the potential [see Figs.~\ref{Fig1}(a)-(b)], confirming an induced bistability in the dynamics. This observation clearly suggests that with increase in the duration of activity, an inertial active Ornstein-Uhlenbeck particle with an internal energy depot displays a bistability in the distribution functions due to an induced additional confinement. Further, in Fig.~\ref{Fig1}(c), we present the steady state velocity distribution $P_{s}^{h}(v)$ for different values of $D$. With increase in $D$ value, the double peak structure in the distribution gets suppressed. At the same time, the distribution becomes wider. This is a clear indication that with increase in the value of noise strength, the effect of additional induced confinement decreases as a result the particle effectively diffuses away inside the confinement.
\begin{figure}
\includegraphics[scale =0.29]{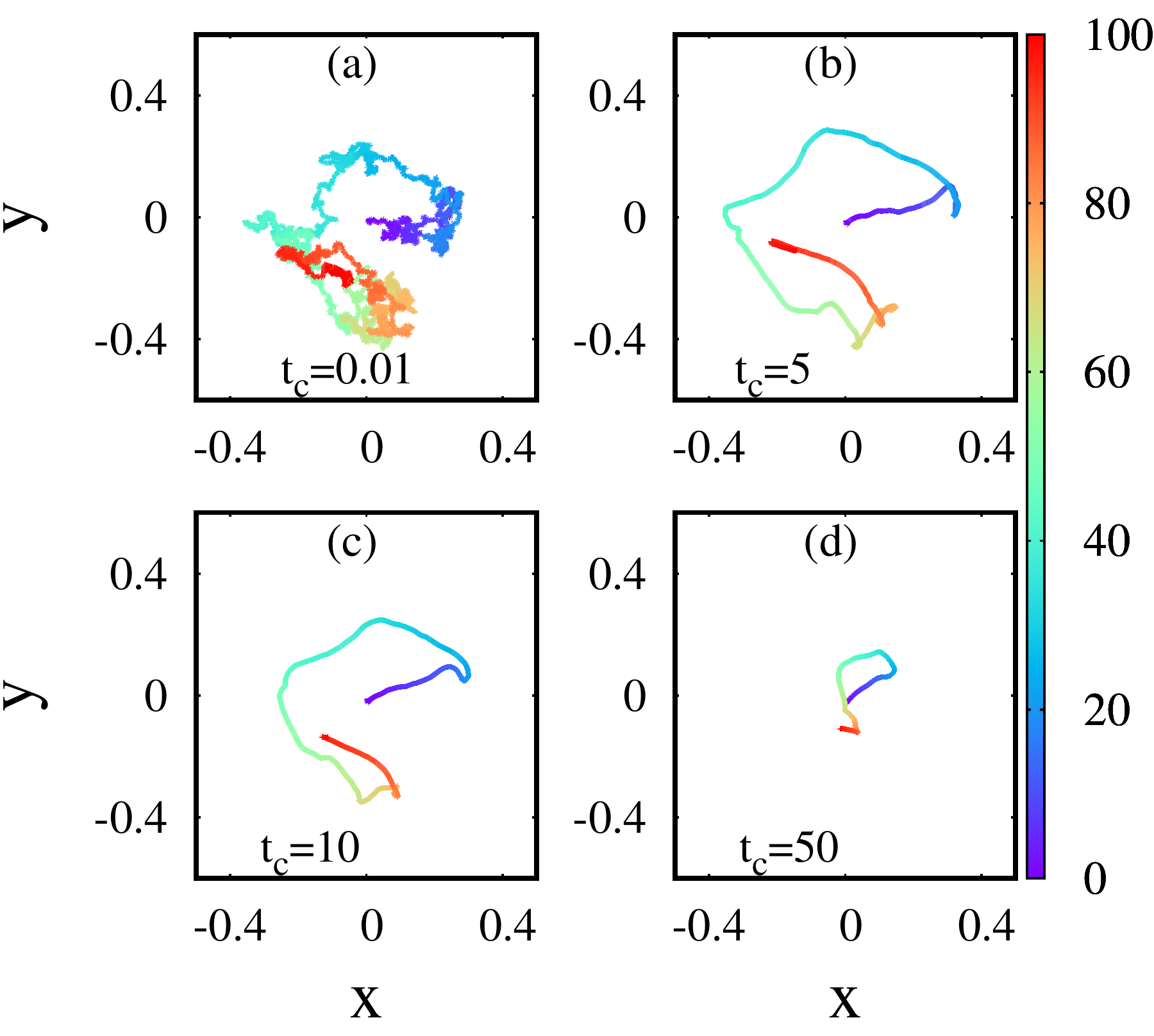}
    \caption{Particle trajectories in $x$-$y$ plane of an overdamped active particle for different values of $t_{c}$. The other common parameters are $\text{D}=k=1,   \text{ q}=3, \text{ d}=0.1, \text{ c}=0.01, \text{ and } \gamma_{0}=50$.}
    \label{Fig2traj}
\end{figure}
However, for a highly viscous medium, where the inertial influence is negligible small, the particle performs random trajectories, evident from the trajectory plot in Fig.~\ref{Fig2traj}(a). With increase in $t_{c}$ value, the trajectories get suppressed [see Figs.~\ref{Fig2traj} (b)-(d)], suggesting as usual trapping or confinement of the particle at the centre of the potential with the persistent duration of activity. The same conclusion can be drawn from the steady state distribution plots for position  $P_{s}^{oh}(x)$ and  velocity $P_{s}^{oh}(v)$ in Fig.~\ref{Fig2}(a) and (b), respectively for different values of $t_{c}$. With increase in $t_{c}$ values, the both $P_{s}^{oh}(x)$ and $P_{s}^{oh}(v)$ becomes narrow and shows sharp peak structure across the centre of the potential. This reflects the strong confinement of the particle for sufficiently long duration of activity.
\begin{figure}
\includegraphics[scale=0.3]{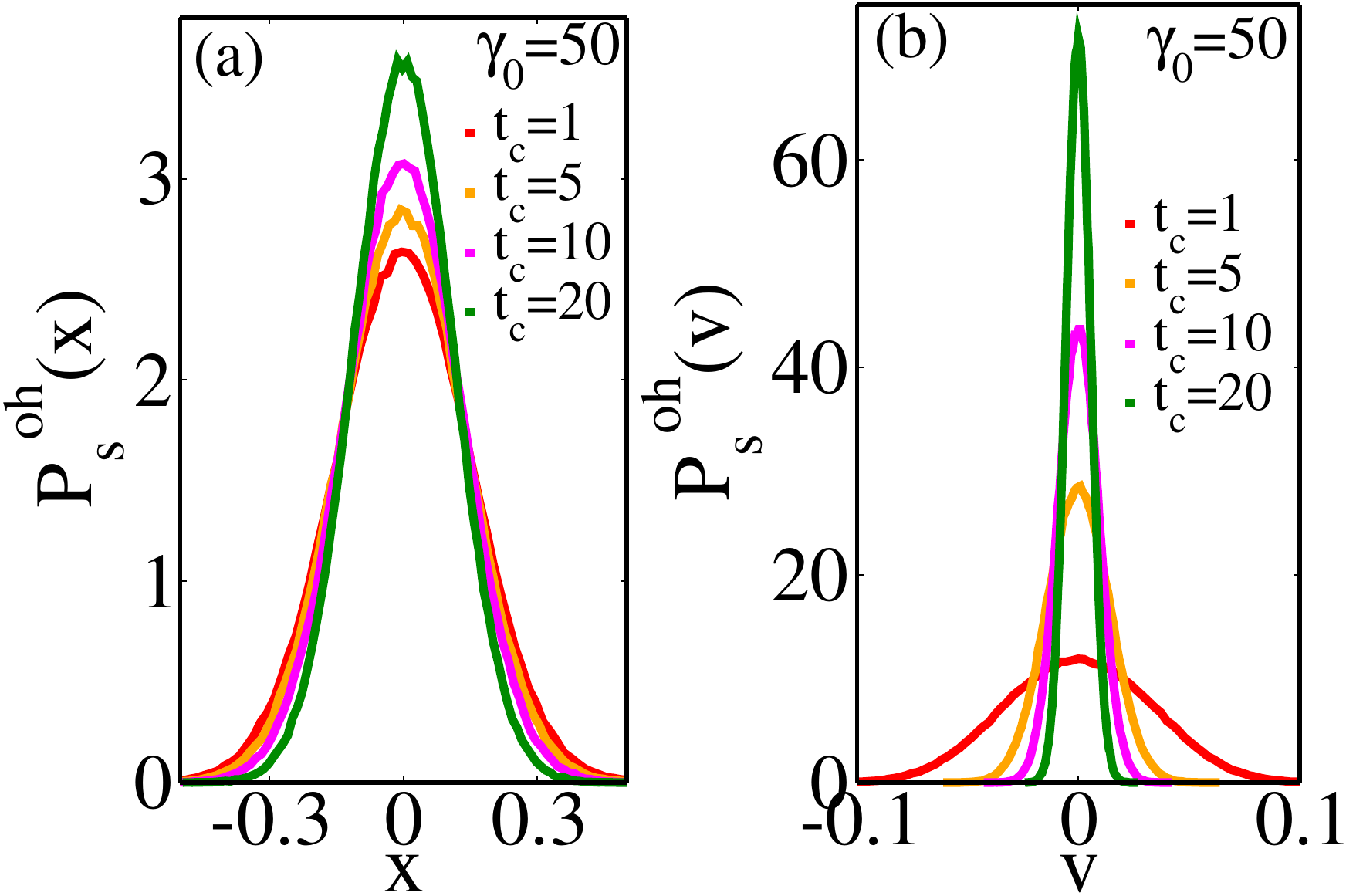}
\caption{$P_{s}^{oh}\left(x\right)$ as a function of $x$ in (a) and velocity distribution $P_{s}^{oh}\left(v\right)$ as a function of $v$ in (b) for different values of  $t_{c}$. The other common parameters are $\text{D}=\text{k}=1,\text{ q}=3,\text{ d}=0.1,\text{ c}=0.01, \text{ and }\gamma_{0}=50$.}
    \label{Fig2}
\end{figure}
Next, we have investigated the distribution functions of position and velocity of the particle in the case when the particle is not confined and set as free $\left[U\left(x\right)=0\right]$. In this case, for the vanishing limit of activity time ($t_{c} \rightarrow 0$), we have obtained the steady state probability distribution for velocity [$P^{f}$] using the perturbative approach as used in Ref.~\cite{martin2021aoup}. The Fokker Plank equation corresponding to the dynamics [Eq.~\eqref{dynamics1}] with $\left[U\left(x\right)=0\right]$ for a free particle is given by
\begin{equation}
    \begin{split}
       \frac{\partial P^{f}(v,\eta,t)}{\partial t} &=\frac{\partial}{\partial v}\left(\gamma_v v-\sqrt{D}\eta(t)\right)P^{f}\\
       &+\frac{\partial}{\partial \eta}\left(\frac{\eta(t)}{t_{c}}+\frac{1}{2t_c^{2}} \frac{\partial}{\partial \eta}\right) P^{f}.
    \end{split}
    \label{fokkerorgfree}
\end{equation}
In the time asymptotic limit or at steady state, the distribution $ P_{s}^{f}(v,\eta, t)$ of Eq.\eqref{fokkerorgfree} satisfies 
\begin{equation}
       LP_{s}^{f}\left(v,\eta\right)=0.
       \label{operfree}
   \end{equation}
   with $L$ as the Fokker-Planck operator and is given by
   \begin{equation}
      L=\frac{\partial}{\partial v}\left(\gamma_{v} v\right)-\eta\left(t\right)\sqrt{D}\frac{\partial}{\partial v}+\frac{1}{t_{c}}\frac{\partial}{\partial \eta}\eta\left(t\right) +\frac{1}{2t_{c}^{2}}\frac{\partial^{2}}{\partial \eta^{2}}.
      \label{operator1free}
   \end{equation} 
In order for expanding the steady state probability distribution function $P_{s}^{f}\left(v,\eta\right)$ in a series form of $t_{c}^{\frac{1}{2}}$, we rescale $\eta\left(t\right)=\dfrac{{\eta\left(t \right)}}{\sqrt{t_{c}}}$. Hence, Eq.~\eqref{operfree} takes the form

   \begin{equation}
     \left[\dfrac{L_{1}}{t_{c}}-\eta\left(t\right) \sqrt{\dfrac{D}{t_{c}}}\dfrac{\partial}{\partial v}+\dfrac{\partial}{\partial v}\left(\gamma_{v}v\right)\right]P_{s}^{f}\left(v,\eta\right)=0.
     \label{operator2free}
   \end{equation}
   
   where, $L_{1}=\left[\dfrac{\partial}{\partial \eta}\eta+\dfrac{1}{2 t_{c}}\dfrac{\partial^{2}}{\partial \eta^{2}}\right]$ is the  operator of Ornstein-Uhlenbeck process. The nth eigen function $P_{n}^{f}\left(\eta\right)$ of the operator $L_{1}$ can be related to the nth Hermite polynomial $H_{n}\left(\eta\right)$ as
  
   \begin{equation}
P_{n}^{f}\left(\eta\right)=\dfrac{e^{-\eta^{2}}H_{n}\left(\eta\right)}{\sqrt{\pi n!2^{n}}},
   \end{equation}
  where $H_{n}\left(\eta\right)=\left(-1\right)^{n} e^{\eta^{2}}\dfrac{\partial}{\partial \eta}^{n}e^{-\eta^{2}}$. The eigen function $P_{n}^{f}\left(\eta\right)$ of the operator $L_{1}$ satisfies the relation
   \begin{equation}
    L_{1}P_{n}^{f}\left(\eta\right)=-n P_{n}^{f}\left(\eta\right).
       \label{l1relationfree}
   \end{equation}
   $P_{n}^{f}\left(\eta\right)$ are further orthogonal to $H_{k}\left(\eta\right)$ and hence as per the orthogonality property, it satisfies  
   \begin{equation}
       \int_{-\infty}^{\infty}\dfrac{H_{k}\left(\eta\right)P_{n}^{f}\left(\eta\right)d\eta}{\sqrt{2^{k}k!}}=\delta_{kn}.
       \label{orthopropertyfree}
   \end{equation}  
   Now, the steady state distribution $P_{s}^{f}\left({v,\eta}\right)$ can be expressed as a series of $A_{n}\text{'s}$ and $P_{n}^{f}\text{'s}$ as 
   \begin{equation}
    P_{s}^{f}\left(v,\eta\right)=\sum_{n}P_{n}^{f}\left(\eta\right)A_{n}\left(v\right).
       \label{soln1free}
   \end{equation}
%The $A_{n}\left(v\right)$ from the above equation can be obtained by multiplying Eq.~\eqref{soln1free} by $\dfrac{H_{k}\left(\eta\right)}{\sqrt{2^{n}n!}}$ and using the orthogonality property from  Eq.~\eqref{orthopropertyfree} as
Using the orthogonality property in Eq.~\eqref{soln1free}, $A_{n}\left(v\right)$ can be obtained as
\begin{equation}
    A_{n}\left(v\right)=\int P_{s}^{f}\left(v,\eta\right) \dfrac{H_{n}\left(\eta\right)}{\sqrt{2^{n}n!}}d\eta .
    \label{anv}
\end{equation}
% To obtain the velocity dependent part$\left[A_{n}\left(v\right)\right]$ of the steady state probability distribution $P_{s}\left(v,\eta\right)$,  Eq.~\eqref{soln1free} can be inserted into Eq.~\eqref{operator2free} and using Eq.~\eqref{l1relationfree}, $A_{n}\left(v\right)$ is a solution of

Now using Eq.~\eqref{soln1free} and Eq.~\eqref{l1relationfree},    Eq.~\eqref{operator2free} can be rewritten as
 \begin{equation}
 \begin{split}
     -\sum_{n}\dfrac{n}{t_{c}}
P_{n}^{f}\left(\eta\right) A_{n}\left(v\right)- \sum_{n}\sqrt{\dfrac{D}{t_{c}}}\eta\left(t\right)P_{n}^{f}\left(\eta\right)\dfrac{\partial}{\partial v}A_{n}\left(v\right)\\ + \sum_{n}P_{n}^{f}\left(\eta\right)\dfrac{\partial}{\partial v} \gamma_{v} v A_{n}\left(v\right)=0 .
\label{priorrecurrence}
\end{split}
\end{equation}

% Using the recurrence property of Hermite polynomial$\left[H_{n+1}\left(\eta\right)=2\eta H_{n}\left(\eta\right)-2nH_{n-1}\left(\eta\right)\right]$, the $\eta\left(t\right) P_{n}\left(\eta\right)$ can be expressed as a sum of $P_{n+1}\left(\eta\right)$ and $P_{n-1}\left(\eta\right)$ as
As per the recurrence property of Hermite polynomial, %$\eta P_{n}^{f}\left(\eta\right)$ can be expressed as a sum of $P_{n+1}^{f}\left(\eta\right)$ and $P_{n-1}^{f}\left(\eta\right)$ 
we can write
  \begin{equation}
      \eta P_{n}^{f}\left(\eta \right)=\sqrt{n+1} P_{n+1}^{f}\left(\eta \right)+\sqrt{n} P_{n-1}^{f}\left(\eta \right) .
      \label{hermiteproperty}
  \end{equation}
%  Substituting Eq.~\eqref{hermiteproperty} into Eq.~\eqref{priorrecurrence} and by using the orthogonality relation from Eq.~\eqref{orthopropertyfree}, we obtain a recurrence relation in terms of the expansion coefficient $A_{n}\left(v\right)$ as

Using Eq.~\eqref{hermiteproperty} and Eq.~\eqref{orthopropertyfree}, Eq.~\eqref{priorrecurrence} can be expressed in terms of a recurrence relation
   \begin{equation}
   \begin{split}
&nA_{n}\left(v\right)=t_{c}\dfrac{\partial}{\partial v}\left(\gamma_{v} v A_{n}\left(v\right)\right)\\
&-\sqrt{Dt_{c}}\left[\sqrt{n+1}\dfrac{\partial}{\partial v}A_{n+1}\left(v\right)+\sqrt{n}\dfrac{\partial}{\partial v}A_{n-1}\left(v\right)\right]. 
     \end{split}
     \label{recurrencefree}
   \end{equation}
%   In order to equate the coefficients of $t_{c}^{\frac{k}{2}}$, the above Eq. can be written as
   
%   \begin{equation}
  % \begin{split}
%nA_{n}^{k}\left(v\right)t_{c}^{\frac{k}{2}}=t_{c}\dfrac{\partial}{\partial v}\left(\gamma_{v} v A_{n}^{k-2}\left(v\right)\right)t_{c}^{\frac{k-2}{2}} \\-\sqrt{Dt_{c}}\left[\sqrt{\left(n+1\right)}\dfrac{\partial}{\partial v}A_{n+1}^{k-1}\left(v\right)+\sqrt{n}\dfrac{\partial}{\partial v}A_{n-1}^{k-1}\left(v\right)\right]t_{c}^\frac{k-1}{2}
%\label{recurrencefree1}
%     \end{split}
%   \end{equation}
Substituting $A_{n}\left(v\right)$ as a series in powers of $t_{c}^{\frac{k}{2}}$ and equating the coefficients of order $t_{c}^{\frac{k}{2}}$ from both the sides of Eq.~\eqref{recurrencefree}, one can obtain
\begin{equation}
\begin{split}
nA_{n}^{k}\left(v\right)&=\dfrac{\partial}{\partial v}\left[\gamma_{v} v A_{n}^{k-2}\left(v\right)\right]-\\&\sqrt{\left(n+1\right)D}\dfrac{\partial}{\partial v}  A_{n+1}^{k-1}\left(v\right) -\sqrt{nD} \dfrac{\partial}{\partial v} A_{n-1}^{k-1}\left(v\right).
    \end{split}
    \label{Ankfree}
\end{equation}
From Eq.~\eqref{soln1free}, the steady state probability distribution for velocity $P_{s}^{f}\left(v\right)$ can be expressed as
   \begin{equation}
       P_{s}^{f}\left(v\right)=\int_{-\infty}^{\infty}P_{s}\left(v,\eta\right)d\eta
    \label{psgeneral}
   \end{equation}
%   and
%   \begin{equation}
%       \int_{-\infty}^{\infty}P_{s}\left(v,\eta\right)d\eta=A_{n}\left(v\right)
%    \label{ps1general}
  % \end{equation}
  Now, in the expansion of $A_{n}(v)$, we consider $A_{n}^{2k}$ as the integer powers of $t_{c}$ and  $A_{n}^{2k+1}$ as the half integer powers of $t_{c}$ and we propose the scaling 
%   Now, the $A_{n}\left(v\right)$'s can be expanded as a series in powers of $t_{c}^\frac{1}{2}$. Here, we consider the $A_{n}^{2k}$ contains only integer powers of $t_{c}$ and  $A_{n}^{2k+1}$ contains only half integer powers of $t_{c}$. Hence, the first non-zero contribution is of $A_{0}$ and order by order in powers of $t_{c}$. We thus propose the scaling 
   \begin{equation}
       A_{0}=A_{0}^{0}\left(v\right)+t_{c} A_{0}^{2}\left(v\right)+t_{c}^{2}A_{0}^{4}\left(v\right)....
   \end{equation}
   For $n=0$, the Eq.~\eqref{anv} can be expressed as
   \begin{equation}
    \int_{-\infty}^{\infty}P_{s}^{f}\left(v,\eta\right)d\eta=A_{0}\left(v\right)=\sum_{k}A_{0}^{2k}t_{c}^{k}
    \label{expansion}
\end{equation}
Using Eq.~\eqref{psgeneral} in Eq.~\eqref{expansion}, $P_{s}\left(v\right)$ can be written as
\begin{equation}
   P_{s}^{f}\left(v\right) =\sum_{k}A_{0}^{2k}t_{c}^{k}.
   \label{sum}
\end{equation}
%We now look for the coefficients $A_{0}^{2k}$. 
For $n=0$, the solution of Eq.~\eqref{Ankfree} can be expressed in terms of $A_{0}$ and $A_{1}$ as 
%  For $n=0$, solution to the Eq.~\eqref{Ankfree} can be expressed in terms of $A_{0}$ and $A_{1}$ as follows:
\begin{equation}
A_{1}^{k-1}\left(v\right)= \sqrt{\dfrac{1}{D}} v\gamma_{v} A_{0}^{k-2}\left(v\right)+b_{k}, 
\label{solnforAfree}
\end{equation}
with $b_{k}$ being the integration constant.
Further, substituting $k=n$ and using  $A_{n}^{j}\left(v\right)=0$ for $j < n$ in Eq.~\eqref{Ankfree}, we get the expression for  $ A_{n}^{n}\left(v\right)$ in terms of $A_{0}^{0}\left(v\right)$
   \begin{equation}
   A_{n}^{n}\left(v\right)=\left(-1\right)^n\sqrt{\dfrac{D}{n!}}\dfrac{\partial}{\partial v}A_{0}^{0}\left(v\right).
   \label{solnAnfree}
\end{equation}
%Here, $A_{0}^{0}\left(v\right)$ corresponds to the stationary probability distribution for $t_{c}=0$. 
Now, substituting $k=2$ in Eq.~\eqref{solnforAfree} and $n=1$ in Eq.~\eqref{solnAnfree}, and equating both these equations, we get  
%The term $A_{0}^{0}\left(v\right)$ can be obtained by substituting $k=2$ in Eq.~\eqref{solnforAfree} and $n=1$ in Eq.~\eqref{solnAnfree} and equating them gives,
\begin{equation}
    \dfrac{\partial}{\partial v}A_{0}^{0}\left(v\right)+ \dfrac{1}{D}\gamma_{v}vA_{0}^{0}\left(v\right)= -\dfrac{b_{2}}{\sqrt{D}} .
   \label{differentialofA0free}
\end{equation}
Fixing the constant of integration $b_{2}$ to zero, the solution of Eq.~\eqref{differentialofA0free} can be approximated as 
%In the above equation, constant $b_{2}$ is fixed to zero. The solution to the above equation can be approximated to be
\begin{equation}
    A_{0}^{0}\left(v\right)=N \left(1+\frac{dv^{2}}{c}\right)^\frac{q}{2D}\exp{\left(\frac{-\gamma_{0}v^{2}}{2D}\right)},
    \label{Aosolnfree}
\end{equation}
 where $N$ is the normalization constant. 
 Similarly, substituting $n=1$ and $k=3$ in Eq.~\eqref{Ankfree}, we get 
  %Now, the next order correction $A_{0}^{2}\left(v\right)$ can be computed as follows. Substituting as $n=1$ and $k=3$ in Eq.~\eqref{Ankfree} gives
 \begin{equation}
 \begin{split}
     A_{1}^{3}\left(v\right)=-\sqrt{2D}\dfrac{\partial}{\partial v} A_{2}^{2}\left(v\right)-\sqrt{D}\dfrac{\partial}{\partial v} A_{0}^{2}\left(v\right)+\\\dfrac{\partial}{\partial v}\left(\gamma_{v}v A_{1}^{1}\left(v\right)\right) 
     \label{A13}
     \end{split}
 \end{equation}
 Now, taking $k=4$ in Eq.~\eqref{solnforAfree}, $n=2$ and $n=1$ in Eq.~\eqref{solnAnfree} and using them in Eq.~\eqref{A13} and solving Eq.~\eqref{A13}, one can get $A_{0}^{2}(v)$. 
Now using $A_{0}^{0}\left(v\right)$ and $A_{0}^{2}\left(v\right)$ in \eqref{sum}, $P_{s}^{f}(v)$ can be obtained upto the second order as
%     Now, substituting the coefficients $A_{0}^{0}$ from Eq.~\eqref{Aosolnfree} and $A_{0}^{2}$ from the solution of Eq.~\eqref{differA02} in Eq.~\eqref{sum} gives the probability distribution for velocity upto the order $t_{c}$. 
\begin{widetext}
\begin{equation}
\begin{split}
P_{s}^{f}\left(v\right)&=Ne^{-\frac{\gamma_{0} v^2}{2 D}} \left(1+\frac{d v^2}{c}\right)^{\frac{q}{2D}}\Bigg\{1+\frac{t_{c}}{D^{5/2}}\Bigg[D \gamma_{0} v \left(D+2 q\right)+\frac{1}{2} \gamma_{0}^{2} v^2 \left(2 D+3 q\right)-\frac{1}{3} D \gamma_{0}^{2} v^3+\frac{c d q^2 \left(q-4 D\right)}{4 \left(c+d v^2\right)^2}\\ &\frac{6 q \left(c \gamma_{0} (4 D-3 q)-d (2 D-q) \left(D v-q\right)\right)}{12 \left(c+d v^2\right)}-\frac{1}{4} \gamma_{0}^3 v^4 -\frac{3 \gamma_{0} q  \left(c\gamma_{0}+d q\right) \log \left(c+d v^2\right)}{2 d} \\&+b_1 \exp -\left(\frac{-d q \log \left(c+d v^2\right)+c \gamma_{0}+d \gamma_{0} v^2}{2 d D}\right)
\Bigg]\Bigg\}.
 \label{probabilityfree}
\end{split}
\end{equation}
 \end{widetext}
  Here $b_{1}$ is the integration constant. In $t_{c}\rightarrow 0$ limit, Eq.~\eqref{probabilityfree} reduces to
  \begin{equation}
P_{s}^{f}\left(v\right)=N\left(1+\frac{dv^{2}}{c}\right)^\frac{q}{2D}\exp{{\left(\frac{-\gamma_{0}v^{2}}{2D}\right)}}
     \label{prblimit}
\end{equation}

  % Here $b_{1}$ is the integration constant. In $t_{c}\rightarrow 0$ limit, Eq.~\eqref{probabilityfree} reduces to
% \begin{equation}
% P_{s}^{f}\left(v\right)=N\left(1+\frac{dv^{2}}{c}\right)^\frac{q}{2D}\exp{{\left(\frac{-\gamma_{0}v^{2}}{2D}\right)}}
%      \label{prblimit}
% \end{equation}
which is same as reported in Ref.~\onlinecite{erdmann2000brownian}. 

\begin{figure}
\centering
  \includegraphics[scale =0.38]{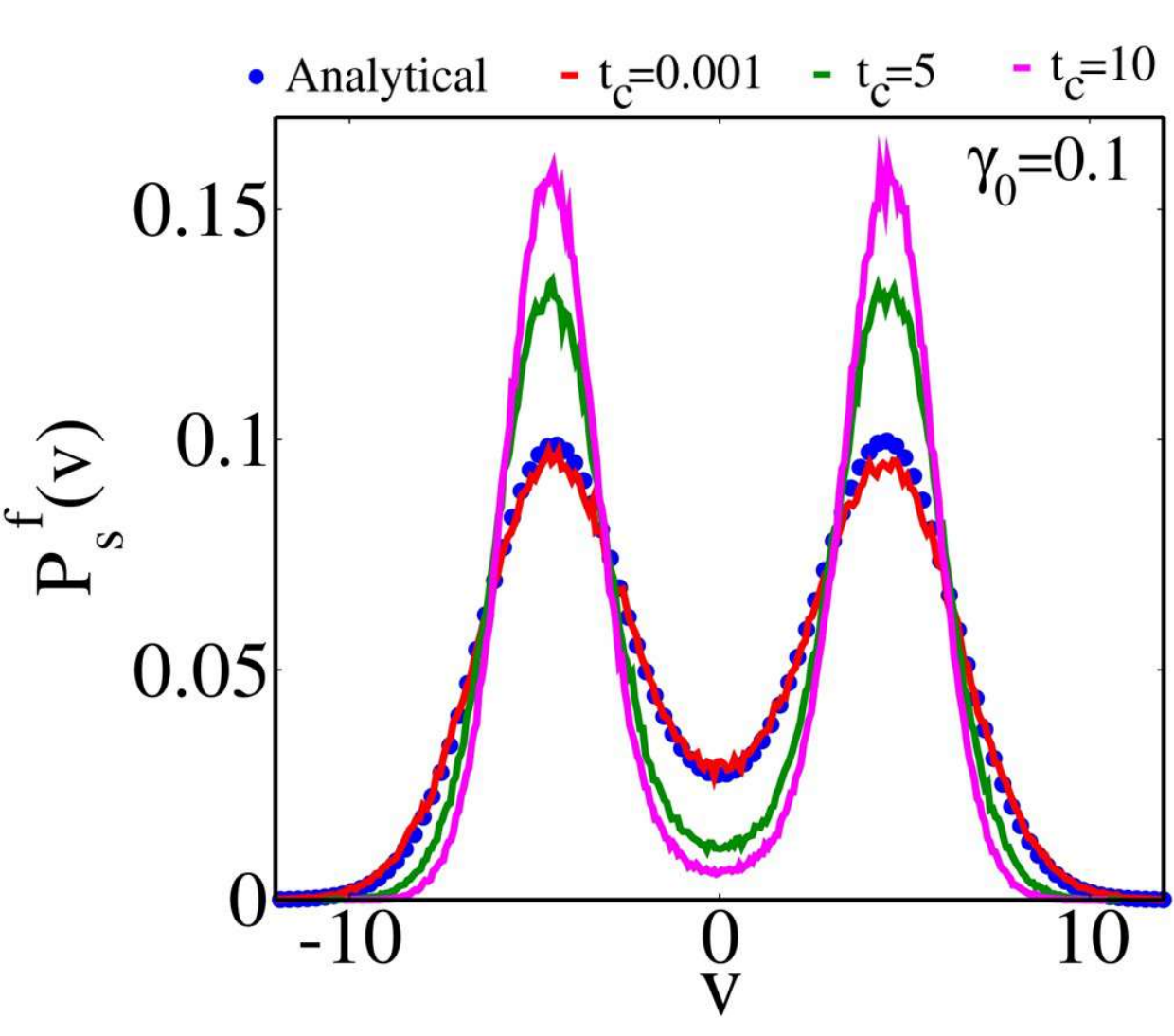}
\caption{$P_{s}^{f}\left(v\right)$ as a function of $v$ for different $t_{c}$ values. The dotted line (blue) corresponds to analytical calculation [Eq.~\eqref{probabilityfree}] and the solid lines correspond to the simulation results. The other common parameters are $ q=3,\text{D}= 1, \text{d}=0.01, \text{ and } \text{c}=\gamma_{0}=0.1.$}
   \label{freesimandana}
\end{figure}

In Fig.~~\ref{freesimandana}, we present the steady state velocity distribution $P_{s}^{f}\left(v\right)$ of a free inertial active particle as a function of $v$ for different values of $t_{c}$. In $t_{c} \rightarrow 0$ limit, the velocity distribution takes a Gaussian shape for either $q=0$ or $d=0$. For a finite $t_{c}$ value, the distribution displays two bell-like peak structure across the centre reflecting a non-Maxwell boltzmann shape same as in case of a confined harmonic well. With further increase in $t_{c}$ value, the distribution transforms into a two separated sharp peak structure across the centre reflecting a kind of bistability in the dynamics. For very low $t_{c}$ value ($t_{c}=0.001$), the analytically calculated velocity distribution function is in good agreement with the simulation results. Moreover, for a highly viscous medium, where the inertial influence is negligible small, the velocity distribution takes as usual Maxwell-Boltzmann shape with a single peak at the centre. These observations clearly suggest that for the dynamics of an inertial active particle in presence of a non-linear friction and with a finite $t_{c}$, there is an additional induced confining mechanism due to which the velocity distribution displays a bimodal or non-Maxwell Boltzmann shape and it transforms into a two separated peak structure with increase in the duration of activity.
This confining mechanism can be explained by the concept of effective temperature of the medium. For a finite $t_{c}$  or finite $q$ value, the system is driven away from equilibrium with an effective temperature not same as the actual temperature of the bath. 

In order to understand the concept of effective temperature, we have computed the steady state mean square velocity $ \langle \Delta v^{2}\rangle=\langle \left(v-\langle v \rangle\right)^{2}$ corresponding to the approximated distribution function $P_{s}^{f}(v)$ [Eq.~\eqref{probabilityfree}], which is given by

\begin{figure}
\includegraphics[scale =0.31]{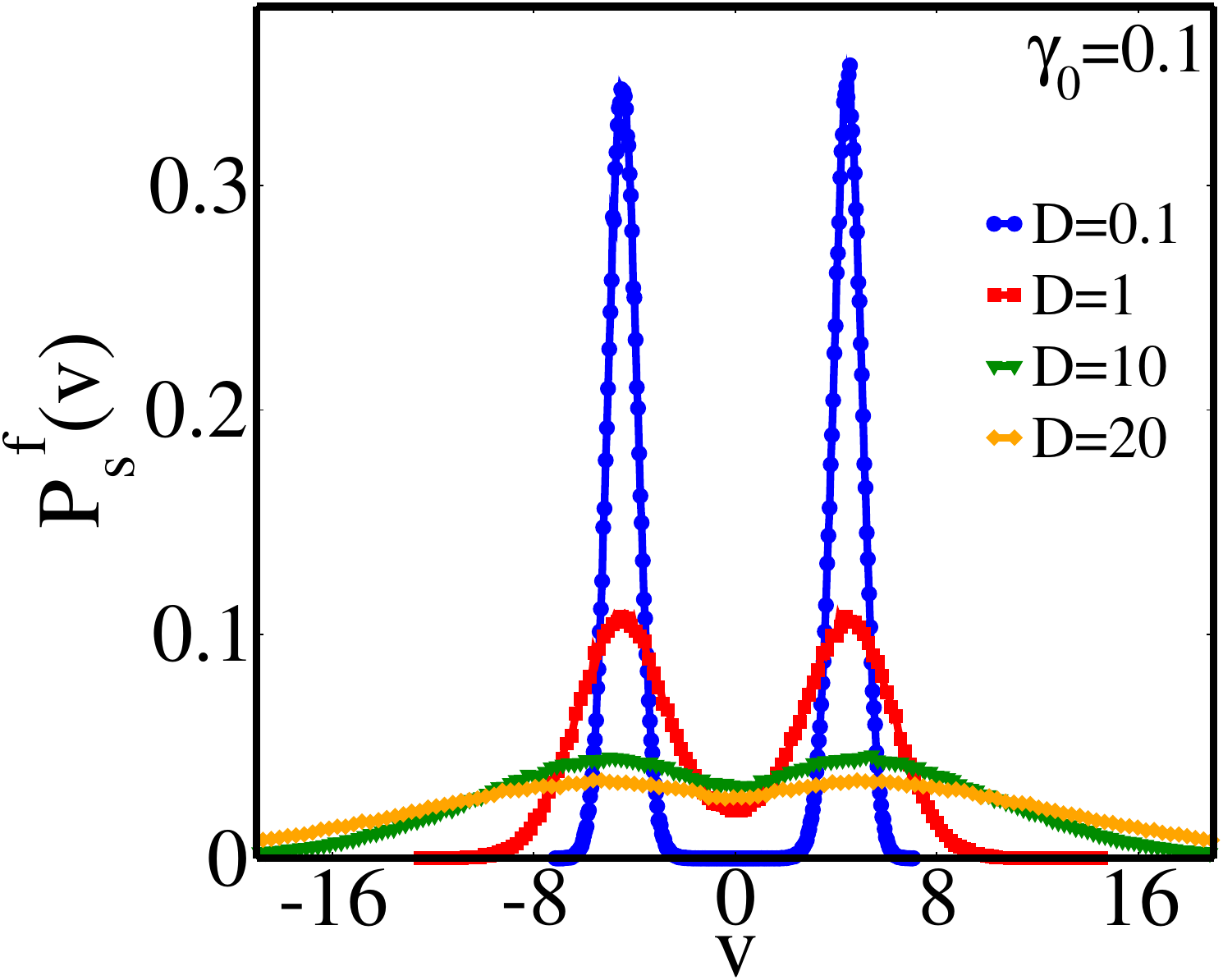}
\caption{$P_{s}^{f}\left(v\right)$ as a function of $v$ for different $D$ values. The other common parameters are $q=3,\text{t}_{c}=1, \text{d}=0.01, \text{ and } \text{c}=\gamma_{0}=0.1$.}
   \label{Dvary}
\end{figure}

\begin{widetext}
\begin{equation}
\begin{split}
   \langle \Delta v^{2}\rangle=&\dfrac{1}{M}\left(\dfrac{D}{\gamma_{0}}\right)^\frac{1}{2}\Bigg\{\left(\dfrac{D}{\gamma_{0}}\right)^{\frac{1}{2}}\left[1+\dfrac{3qd}{2c\gamma_{0}}\right]+t_{c}\Bigg[-\dfrac{qc\gamma_{0}}{2dD^{\frac{3}{2}}}\left(1+\dfrac{q}{(2D-d)}\right)\\&+\dfrac{1}{d\sqrt{D}}\left[c\gamma_{0}+q\left(\dfrac{2c\gamma_{0}}{2D-q}-d\right)-q^{2}\left(\dfrac{7d}{2(2D-q)}+\dfrac{c\gamma_{0}}{2\left(8D^{2}-6Dq+q^{2}\right)}\right)-q^{3}\left(\dfrac{3d}{2\left(8D^{2}-6Dq+q^{2}\right)}\right)\right]\\&+\sqrt{D}\Bigg[\dfrac{2c\gamma_{0}}{d}\left(\dfrac{1}{(2D-q)}+\dfrac{c\gamma_{0}}{d\left(8D^{2}-6Dq +q^{2}\right)}\right)+q\left(\dfrac{1}{2D-q}-\dfrac{4c^{2}\gamma_{0}^{2}}{d^{2}\left((2D-q)(4D-q)(6D-q)\right)+q^{2}}\right)\\&+q^{2}\left(\dfrac{15d}{2(4D-q)c\gamma_{0}}-\dfrac{6d}{c\gamma_{0}(2D-q)}+\dfrac{15}{2\left(24D^{2}-10Dq+q^{2}\right)}\right)\Bigg]\\&+\left(\dfrac{D^{3}}{\gamma_{0}}\right)^{\frac{1}{2}}\left[3\sqrt{\gamma_{0}}\left(\dfrac{2}{(2D-q)}-\dfrac{5}{(4D-q)}\right)+q\dfrac{3d}{c\sqrt{\gamma_{0}}}\left(\dfrac{15}{4D-q}-\dfrac{2\sqrt{\gamma_{0}}}{2D-q}-\dfrac{35}{2(6D-q)}\right)\right]\Bigg]\Bigg\}.
   \label{vsquare}
    \end{split}
\end{equation}
\end{widetext}
For the system effectively to be at equilibrium, imposing the stationarity of the process and extending the concept of equipartition theorem, one can evaluate the approximate expression for the effective temperature, which is directly related to the strength of noise $D$. In $t_{c}\rightarrow 0$ limit, $\langle \Delta v^{2}\rangle$ in Eq.~\eqref{vsquare} reduces to
    \begin{equation}
   \langle \Delta v^{2}\rangle=\dfrac{\left(\frac{\text{D}}{\gamma_{0}}\right)\left[1+\frac{3 d q}{2 c\gamma_{0}} \right]}{1+\dfrac{qd}{2c\gamma_{0}}}.
\end{equation}
In this limit, the effective temperature can be calculated as $\frac{\langle \Delta v^{2}\rangle}{k_{B}}$ and hence it is linearly related to $D$. For $D=2 \gamma_{0} k_{B} T$ along with $q=0$, the system approaches thermal equilibrium with $T$ as the temperature of the bath and $k_{B}$ as the Boltzmann constant. In order to understand the impact of noise strength $D$ on the velocity distribution, we have plotted $P_{s}^{f}\left(v\right)$ as a function of $v$ for different values of $D$ in Fig.~\ref{Dvary}. For a non-zero finite $D$ value, $P_{s}^{f}\left(v\right)$ shows as usual non-Gaussian form of distribution with two separated sharp peaks across the centre. This might be due to the fact that there is an induced confinement for the inertial motion of an active Ornstein-Uhlenbeck particle in a non-linear friction. Further, with increase in $D$ value, the peaks of the distribution get suppressed. This is because, with increase in $D$ value, the effective kinetic temperature of the medium increases as it is linearly related to $D$. As a consequence, the particle becomes more energetic and starts diffusing. That is why, with increase in $D$ value, the two peaks in the distribution gets suppressed and the distribution gets broadened. This observation clearly indicates that the effect of induced confining mechanism suppresses with increase in strength of noise and the particle diffuses away, resulting a broadened distribution.

\section{Conclusion}
In this work, we have studied the dynamics of an Ornstein-Uhlenbeck particle in the presence of a non-linear velocity dependent friction force. From the simulation result of position and velocity distribution functions, we observe that when an inertial active particle is confined in a harmonic well, there is an induced bistability with increase in persistent duration of activity. As a consequence the particle prefers to accumulate near the boundary of the potential instead of mean position. On the contrary, for a highly viscous medium, where the inertial influence is negligible small, the distribution shows a single peak structure at the centre of the potential, affirming as usual confinement of the particle. Interestingly, the similar attributes are observed even when the particle is set free. This observation clearly suggests that an induced effective confining mechanism appears for the inertial active motion of a free particle, whose effect subsides with increase in the effective temperature of the medium or strength of noise. Moreover, using perturbative method, we have analytically computed the velocity  distribution function of a free particle in the vanishing limit of noise and the distribution agrees well with the simulation results for low $t_{c}$ values. Finally, we believe that the findings of our model are applicable for controlling the steady state properties of active particle, for instance controlling the effective kinetic temperature of the particle by tuning the persistence duration of activity or noise strength of the medium.

\section{Acknowledgments}
We thank the 8th statphysics community meeting (ICTS/ISPCM2023/02), during which a part of the work was undertaken. 
MS acknowledges the start-up grant from UGC, state plan fund from university of Kerala, SERB-SURE grant (SUR/2022/000377), CRG grant (CRG/2023/002026) from DST, Govt. of India for financial support. AN thanks Muhsin Muhammed for the numerical help.
  \appendix
  \section{}
   \begin{widetext}
   \begin{equation*}
    \begin{split}
        M=&\left(\dfrac{D}{\gamma_{0}}\right)^{\frac{1}{2}}\left[1+\dfrac{dq}{2c\gamma_{0}}\right]+t_{c}\Bigg\{\dfrac{c\gamma_{0}^{\frac{3}{2}}}{2d D^{2}}\left[q-q^{2}\left(\dfrac{1}{2D-q}\right)\right]+\dfrac{\sqrt{\gamma_{0}}}{D}\Bigg[\dfrac{c}{d}-q-q^{2}\left(\dfrac{5}{2(2D-q)}-\dfrac{c\gamma_{0}}{d\left(8D^{2}-6Dq+q^{2}\right)}\right)\Bigg]+\\&\sqrt{\gamma_{0}}\Bigg[\left(\dfrac{2c\gamma_{0}}{d(2D-q)}\right)+q\left(\dfrac{c\gamma_{0}-d}{2D-q}-\dfrac{5c\gamma_{0}}{d\left(8D^2-6Dq+q^{2}\right)}-\frac{4  c^2  \gamma_{0}^{2} }{d^2 (2D-q) (4D-q) (6D-q)}\right)\\&+q^{2}\left(\dfrac{3d}{2(4D-q)c\gamma_{0}}-\dfrac{4d}{(2D-q)\gamma_{0}}+\dfrac{3}{2(24D^{2}-10Dq+q^{2})}-\dfrac{(3d+c\gamma_{0})}{d(8 D^2-6 Dq+q^2)}\right)\Bigg]+\\&D\sqrt{\gamma_{0}}\left[\dfrac{4}{(2D-q)}-\dfrac{6}{(4D-q)}+\dfrac{qd}{c\gamma_{0}}\left(\dfrac{9}{4D-q}-\dfrac{2}{2D-q}-\dfrac{15}{2(6D-q)}\right)\dfrac{3\gamma_{0}}{4D-q}\right]\Bigg\}
        \end{split}
    \end{equation*}
    \end{widetext}

\end{document}